\documentclass[twocolumn,prb,showpacs,secnumarabic,nobibnotes]{revtex4}%
\usepackage{amsfonts}
\usepackage{amsmath}
\usepackage{amssymb}
\usepackage{graphicx}

\newcommand{\Bpar}{B_{||}}
\newcommand{\cond}{e^{2}/h}
\newcommand{\Dpp}{\Delta_{\mathrm{pp}}}
\newcommand{\gmax}{G_{\mathrm{max}}}
\newcommand{\kB}{k_{\mathrm{B}}}

\newcommand{\TK}{T_{\mathrm{K}}}

\newcommand{\Vg}{V_{g}}
\newcommand{\Vdc}{V_{\mathrm{dc}}}
\newcommand{\uV}{\mu \mathrm{V}}

\begin{document}
\title{Zero-bias Anomaly of Quantum Point Contacts in the Low-Conductance Limit}
\author{Y. Ren, W. W. Yu}
\author{S. M. Frolov}
\altaffiliation{Present address: Kavli Institute of Nanoscience, Delft University of Technology, 2628 CJ Delft, The Netherlands.}
\author{J. A. Folk}
\affiliation{Department of Physics and Astronomy, University of British Columbia, Vancouver, BC V6T 1Z1, Canada}
\author{W. Wegscheider}
\affiliation{Laboratorium f\"{u}r Festk\"{o}rperphysik, ETH Z\"{u}rich, 8093 Z\"{u}rich, Switzerland}
\begin{abstract}
Most quantum point contacts (QPCs) fabricated in high-mobility 2D electron gases show a zero-bias conductance peak near pinchoff, but the origin of this peak remains a mystery.   Previous experiments have primarily focused on the zero-bias peak at moderate conductance, in the range $1-2\cond$. Here, measurements are presented of zero-bias peaks that persist down to $10^{-4}\cond$. Magnetic field and temperature dependencies of the zero-bias peak in the low-conductance limit are qualitatively different from the analogous phenomenology at higher conductance, with implications for existing theoretical models of transport in low-density QPCs.

\end{abstract}
\date{\today}
\maketitle

\section{Introduction}

Quantum point contacts (QPCs) are short one-dimensional constrictions, typically fabricated in clean semiconductor 2D electron gases, that have been drawing attention in the condensed matter community for over two decades.  QPCs are an integral part of nearly every 2DEG nanostructure, from quantum dots to Aharanov-Bohm rings, and in principle they are one of the easiest mesoscopic systems to analyze from a theoretical point of view.  On one level, QPCs seem to follow a straightforward single-particle description. Differential conductance at low magnetic field is quantized as $G\equiv dI/dV = N\times2\cond$, where $N$ counts the  spin-degenerate one dimensional subbands in the constriction.\cite{WeesPRL88, WharamJPC88}   QPC conductance on the plateaus is  robust against the effect of interactions: a low-temperature suppression of conductance that might be expected due to Luttinger liquid physics\cite{KanePRB92} disappears when connected to non-interacting leads.\cite{TaruchaSSP95, MaslovPRB95}

But experiments have revealed two characteristic deviations from the non-interacting picture that are observed in most QPCs fabricated in a high mobility material.  First, a shoulder-like feature appears in the linear conductance around $0.7\times2\cond$, which is therefore referred to as ``0.7 structure''.\cite{ThomasPRL96}   Second, a narrow zero-bias peak (ZBP) is observed in source-drain conductance for low magnetic fields.\cite{CronenwettPRL02}  It is generally believed that these conductance features arise from electron-electron interactions.  Many explanations have been proposed, including spontaneous spin polarization,\cite{ThomasPRL96, BerggrenPRB96, ReillyPRL02} 1D Wigner crystallization,\cite{MatveevPRL04} electron-phonon scattering\cite{MatveevPRL03} and Kondo screening of a quasi-localized state.\cite{CronenwettPRL02, MeirPRL02, RejecNature06, Ensslin09} But there remains no consensus on which interpretation is correct, and the subject is still widely debated.

Most previous measurements of ZBPs in point contacts have focused on the high-conductance regime ($G\gtrsim1\cond$), where the 0.7 structure is observed. At high conductance, the ZBP is often attributed to Kondo effect screening of an impurity that is formed self-consistently in the QPC.\cite{CronenwettPRL02, RejecNature06}  Point contact ZBPs typically persist down to much lower conductance,\cite{SarkozyPRB09} and it is tempting to attribute the same mechanism to the formation of ZBPs through the full range $0<G<2\cond$.  As pointed out in Refs.~\onlinecite{CronenwettPRL02} and \onlinecite{SarkozyPRB09}, however, there are several quantitative differences between high conductance ZBPs and those below $G\sim1\cond$.  For example,  ZBPs in both low- and high-conductance regimes split with in-plane magnetic field, but the magnitude of the splitting is very different well above and well below $G\sim1\cond$.\cite{SarkozyPRB09}


In this paper, we report the magnetic field and temperature dependencies of ZBPs from defect-free quantum point contacts deep in the tunneling regime.  Whereas Ref.~\onlinecite{SarkozyPRB09} studied ZBPs covering a wide range of conductances, this work focuses specifically on ZBP phenomenology below $G\sim0.1\cond$. So close to pinch-off, low-conductance ZBP phenomenology can be clearly distinguished from high-conductance behaviors, and provides a test for theoretical models in a new regime.  For example, ZBPs at high conductance include a single-particle contribution due simply to tunneling through a saddle-point barrier,\cite{GlazmanJPCM89, ButtikerPRB90, MartinMorenoJPCM92} which must be taken into account before attempting to discern many-particle physics from this feature. ZBPs at low conductance, on the other hand, are absent in the single-particle picture,\cite{MartinMorenoJPCM92} making them ideal for studying the many-body effects in low-density QPCs. The paper is organized as follows:  Section II presents measurements of ZBPs down to $10^{-4}\cond$ at base temperature and zero field.  Magnetic field and temperature dependencies are presented in Sec. III and Sec. IV respectively, and compared with those of high-conductance ZBPs. In Sec. V, implications of the measurement results on theoretical models are discussed. A brief conclusion is given in Sec. VI.

\begin{figure}[t]
\includegraphics[scale=1]{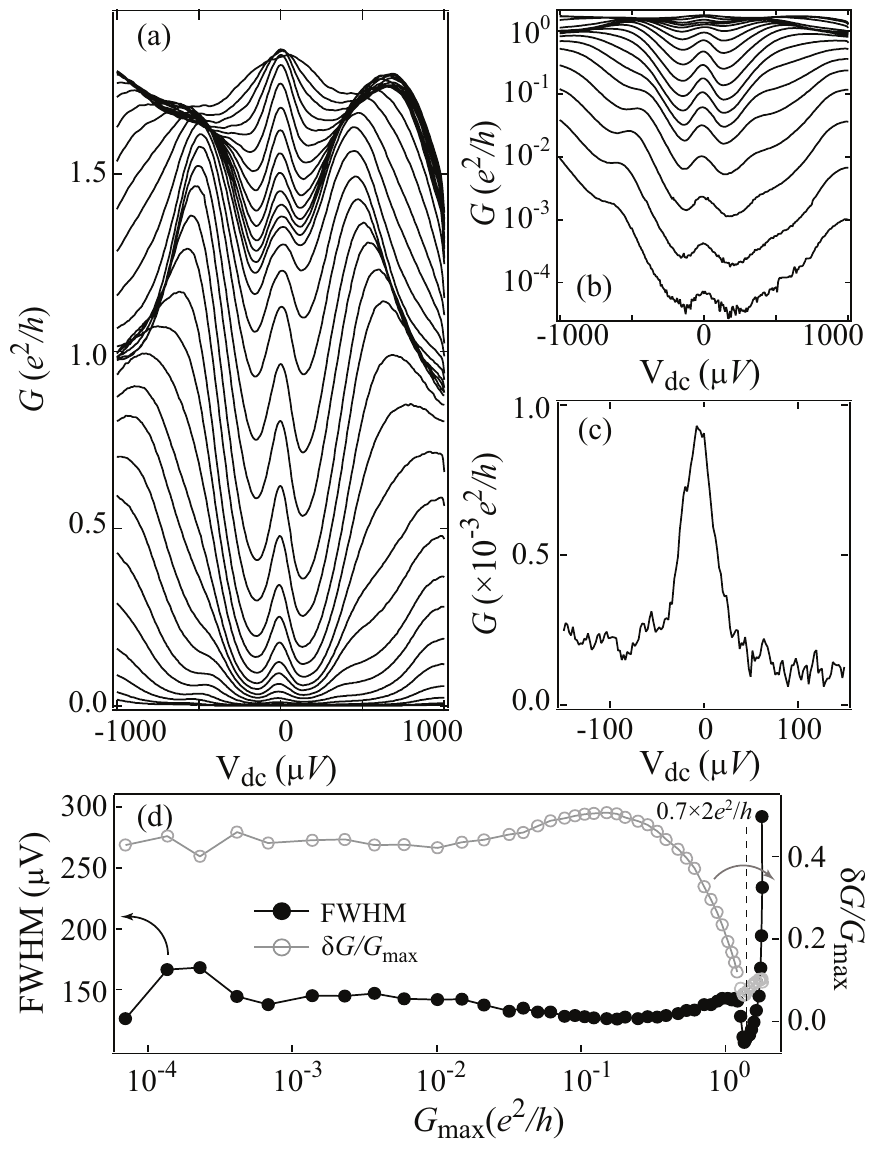}\\
\caption{(\textbf{a}) Differential conductance versus source-drain d.c.~bias, $\Vdc$.  Each trace represents a different gate voltage, $\Vg$, evenly spaced with intervals of 1mV, at $T=40$mK and $\Bpar=0$T.
(\textbf{b}) Logarithmic plot for data in (a). The ZBP was clearly resolved down to $10^{-4}\cond$.
(\textbf{c}) An example of sharp ZBP observed below $10^ {-3}\cond$.
(\textbf{d}) FWHM (left axis) and the relative peak height $\delta G/\gmax$ (right axis) of ZBPs versus conductance maximum, $\gmax$, extracted from data shown in panels (a) and (b). Both FWHM and $\delta G/\gmax$ show a minimum around $0.7\times 2\cond$ and remain basically flat in the low-conductance regime.}
\label{}\end{figure}

\section{Zero-bias Peaks at Low Conductance}

Three $1\mu$m-long and six $0.5\mu$m-long QPCs were defined by electrostatic gates on a GaAs/AlGaAs heterostructure with a two-dimensional electron gas (2DEG) 110 nm below the surface.  The lithographic width of the QPCs was 225 nm.  At $T=1.5 $K, the electron density and mobility of the 2DEG were $n_s=1.11\times 10^{11}$cm$^{-2}$ and $\mu=4.44\times10^{6}\mathrm{cm}^{2}/ \mathrm{Vs}$, respectively. Differential conductance measurements were performed in a dilution refrigerator with base electron temperature $\sim40$mK, using an a.c.~lock-in technique with $V_{\mathrm{ac}}=10 \uV$. Data presented in this paper were measured over a range of gate voltages $\Vg$, source-drain d.c.~bias $\Vdc$, temperature $T$ and in-plane magnetic field $\Bpar$.  For some cooldowns the in-plane field was aligned along the QPC axis, and other times perpendicular to it, but no consistent effect of field orientation was observed.  The data for the figures in this paper came from three different devices; consistent behaviors were observed in all nine devices, independent of length.

Differential conductance signatures of the QPCs in this experiment were similar to those reported across the literature (Fig.~1a).\cite{CronenwettPRL02, SarkozyPRB09} ZBPs were observed from just below the first plateau ($G=2\cond$) all the way down to pinch-off. A logarithmic plot of $G$ (Fig.~1b) shows that the ZBPs could be resolved down to $10^{-4}\cond$, the lowest conductance measured in this experiment. At this low conductance, the detailed shape of differential conductance curve was found to differ between QPCs and even cooldowns.  In many cases the ZBP was very sharp, with peak conductance as large as five times higher than the conductance off-peak even below $G\sim10^{-3}\cond$ (Fig.~1c). The strength and visibility of the ZBP below $0.1\cond$ depended on device details, but all measured QPCs showed ZBPs at least down to $10^ {-1}\cond$.  We conclude, therefore, that the presence of a ZBP in the low-conductance limit is a universal characteristic.

ZBPs can be characterized by a full width at half maximum (FWHM) and a peak height, $\delta G$.  Previous reports have consistently shown that the FWHM decreases monotonically as $G$ drops from $2\cond$ to  $\sim 0.7\times2\cond$.\cite{CronenwettPRL02}  Below this conductance, there is a sharp rise in FWHM, which then remains constant down to pinch-off.\cite{CronenwettPRL02, SarkozyPRB09, ChouAPL05} This behavior is clearly seen in Figs.~1b and 1d.  Low conductance ZBPs from the devices in this experiment had FWHMs within the range $80-200 \uV$.

The peak height, $\delta G$ , can be defined as the difference between the conductance on top of the peak, $\gmax$, and the average of local minima on either side.  Existing literature describes a similar non-monotonic dependence of $\delta G$ on $\gmax$, with a local minimum at $G\sim0.7\times2\cond$.\cite{SarkozyPRB09} One significant feature of $\delta G$ that can be easily seen in log-scale plots such as Fig.~1b, but to our knowledge has not been previously pointed out, is that the {\em relative} peak height $\delta G/\gmax$ saturates at a value that does not change over orders of magnitude in conductance (see also Fig.~1d). This remarkable consistency of relative peak height over a wide range of conductance was observed in many QPCs and cooldowns. The saturation value varied within the range $0.3-0.8$ from device to device.  A similar saturation of $\delta G/\gmax$ can be noted in Fig.~2b of Ref.~\onlinecite{SarkozyPRB09} but was not discussed in that work.

\begin{figure}[t]
\includegraphics[scale=1]{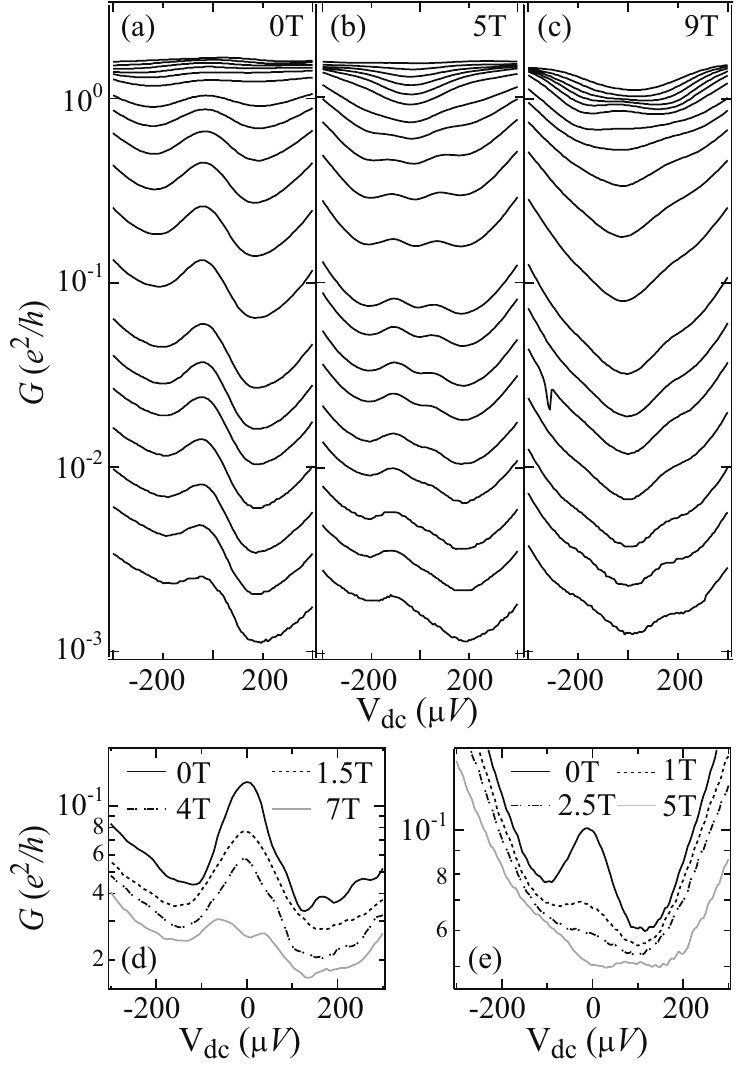}\\
\caption{Evolution of the low-conductance ZBPs in an in-plane field (\textbf{a}) $\Bpar=0$T, (\textbf{b}) $\Bpar=5$T, and (\textbf{c}) $\Bpar=9$T. Individual traces in (a-c) represent evenly-spaced gate voltages as in Figs.~1ab, with every other trace removed above $G=1\cond$ for clarity.
(\textbf{d}) An example of a ZBP with splitting much less than $2E_z$.
(\textbf{e}) An example of a ZBP that collapses before clear splitting is observed.
In panels (d) and (e), gate voltages are different for different fields, chosen to maintain the conductance at $\Vdc=-400\uV$. Traces for $\Bpar>0$T are offset vertically for clarity.
}\label{} \end{figure}

\section{Splitting in Magnetic Field}
Zero-bias conductance peaks in QPCs are suppressed by in-plane magnetic fields on the scale of several Tesla---a phenomenon observed in this experiment and consistent with reports from across the literature.  For ZBPs above $1\cond$, a splitting was often observed before the peak was fully suppressed. The magnitude of the splitting, $\Dpp$, was typically between 3-5$E_z$,\cite{CronenwettPRL02, SarkozyPRB09} where $E_z=|g\mu_B \Bpar|$ is the Zeeman energy using the bulk GaAs g-factor, $g=-0.44$. We compare $\Dpp$ to Zeeman rather than orbital energy scales because the magnetic fields were applied in the sample plane, causing relatively minor orbital effects.  As the gate voltage was tuned to bring $G$ below $1\cond$ the splitting in all devices dropped to less than $2E_z$, consistent with the gate voltage dependence of the splitting reported in Ref.~\onlinecite{SarkozyPRB09}.  For even lower conductances,  however, $\Dpp$ saturated to a value that did not change down to pinchoff (see, e.g., Fig.~2b).\cite{Liu09}

The detailed magnetic field dependence of the ZBPs below $G\sim1\cond$ varied widely from device to device, even for lithographically-identical QPCs free of  disorder (resonances).  These diverse behaviors may help explain the range of reports that have appeared in the literature.\cite{CronenwettPRL02, SarkozyPRB09, ChenPRB09, Liu09}  Figure 2 summarizes the magnetic field dependencies that were observed  in this experiment, all for ZBPs with similar zero-field widths and heights.  In Fig.~2b, a clear splitting is observed at intermediate field, and the magnitude saturates to  $1.8E_z$ in the low-conductance regime.  In Fig.~2d, a splitting is again easily seen at $7T$, but the magnitude is only $0.5E_z$.  In Fig.~2e, the peak collapses much more rapidly with field, reducing in height by 66\% at 1T compared to 15\% at 1.5T in Fig.~2d. Small bumps consistent with remnants of a split ZBP are visible at 5T, but the visibility of these features is qualitatively worse than in the other two devices.  As seen in these examples, the magnitude of splitting and resilience of the ZBP in a finite magnetic field are not clearly correlated.

Some of the factors that influence ZBP splitting were explored in Ref.~\onlinecite{ChenPRB09}, where a transition  from splitting to non-splitting behavior was reported by laterally shifting the QPC. In the present experiments, it was observed that ZBP splitting in some QPCs changed from hour to hour with other parameters held unchanged, apparently due to minor rearrangement of dopant potentials that were too small to affect the zero-field conductance. This observation indicates that ZBP splitting is exquisitely sensitive to the energy profile of the QPC, in contrast to the ZBP itself, which was observed in every QPC measured and whose shape was significantly less sensitive to fine details in the QPC potential.

\begin{figure}[t]
\includegraphics[scale=1]{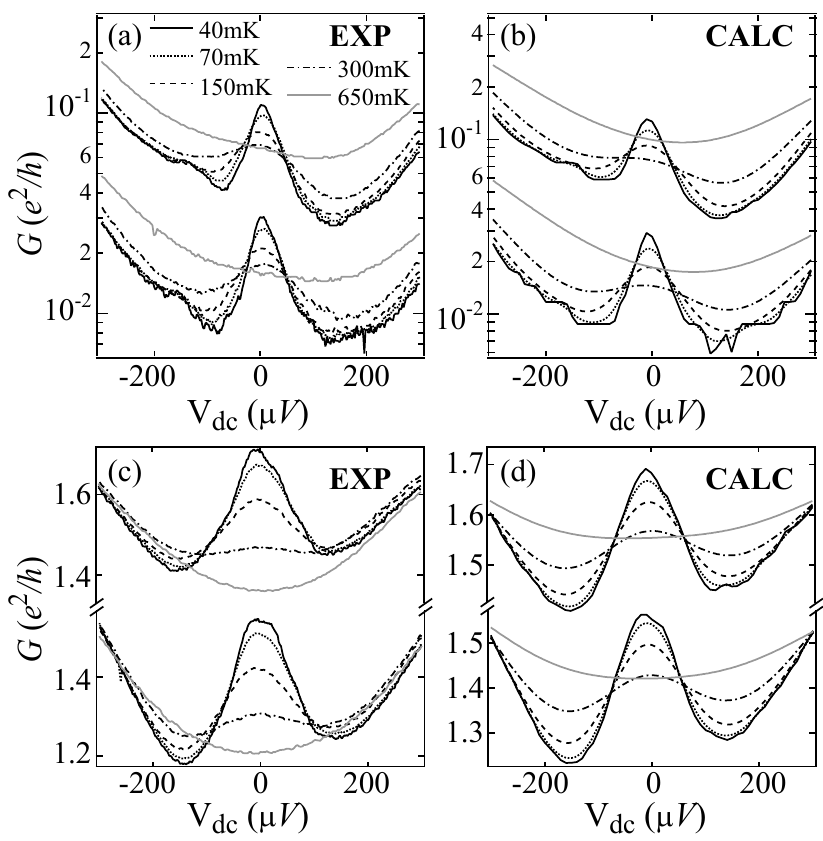}\\
\caption{Evolution of the ZBPs with temperature from $T=40$mK to 650mK.
(\textbf{a, c}) Experimental data for ZBPs in the low-conductance (a) and high-conductance regimes (c).
(\textbf{b, d}) Calculation results for ZBPs in the low-conductance (b) and high-conductance regimes (d).
For each group of curves in (b) and (d), ZBPs at $T\geq 70$mK were approximated by substituting the measured $G(V,T$=40mK) for $G(V,T=0)$ in Eq.~(\ref{simulation}).
}\label{}
\end{figure}

\section{Suppression at Finite Temperature}
As temperature is increased, ZBPs become lower and eventually disappear, independent of conductance  (Figs.~3a, c).\cite{CronenwettPRL02} Using a Landauer description of ballistic transport at bias voltage, $V$, and temperature, $T$,\cite{BeenakkerSSP91}
\begin{equation}\label{Landauer}
      I=\int^{\infty}_{-\infty}dE~t(E,V,T)[f(E,T)-f(E+eV,T)],
\end{equation}
it is seen that the suppression of ZBPs at high temperature can result from broadening of the Fermi functions, $f(E,T)$, or from temperature-dependent changes in the transmission coefficient $t(E,V,T)$, or both. To distinguish these effects, experimental data up to $650$mK are compared to calculations that include thermal broadening but exclude any temperature dependence of $t$ (Fig.~3). Assuming that the voltage bias drops equally on both sides of the QPC\cite{MartinMorenoJPCM92, PatelPRB91} and that the barrier itself is not directly affected by the applied bias, differential conductance at finite temperature $G(V,T)$ can be expressed as the convolution of its zero-temperature value $G(V,T=0)$ with the derivative of the Fermi function:
\begin{equation}
\label{simulation}
     G(V,T)=\int^{\infty}_{-\infty}dV'G(V',0)\frac{\partial f\left(\mu+{(V-V')\over 2},T\right)}{\partial V'},
\end{equation}
where $\mu$ is the chemical potential.

In the low-conductance regime, the simulation results (Fig.~3b) closely resemble the experimental data (Fig.~3a), indicating that thermal broadening due to $f(E,T)$ is the major contributor to the suppression of ZBPs. In the high-conductance regime, however, the zero-bias conductance in the measurement (Fig.~3c) is substantially lower than in the calculation (Fig.~3d), and cannot be accounted for simply by thermal broadening. This suggests that the functional form of $t(E)$ is directly affected by temperature  at high conductance, but not at low conductance.

\begin{figure}[t]
\includegraphics[scale=1]{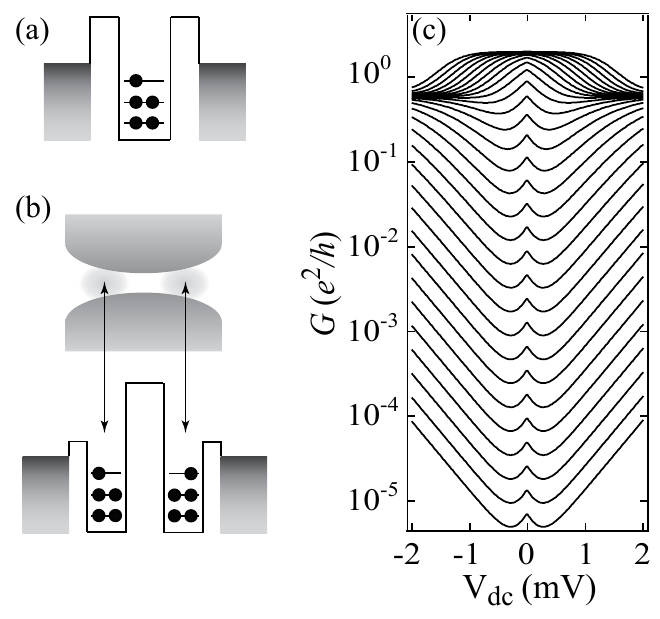}\\
\caption{(\textbf{a}) Energy profile of a symmetric Kondo-like localized state.
(\textbf{b}) Top: geometry of a symmetric QPC with two asymmetric Kondo-like localized states on either end.
Bottom: energy profile of this QPC. Arrows indicate the position of localized states.
(\textbf{c}) Calculated differential conductance versus d.c.~bias using the saddle-point model.\cite{ButtikerPRB90, MartinMorenoJPCM92} The subband energy was assumed to rise linearly with $\Vdc$ at a rate of 0.2meV/mV.\cite{ChenPRB09} Other parameters were adopted from Ref.~\onlinecite{PatelPRB91R}.
}\label{}
\end{figure}

\section{Comparison with theoretical models}

The ZBP is frequently discussed in connection with a possible Kondo effect in QPCs.\cite{CronenwettPRL02, RejecNature06}  Self-consistent density functional calculations suggest that a quasi-localized state may exist in the middle of a low-density 1D constriction (Fig.~4a).\cite{RejecNature06}  Kondo screening of this state would lead to enhanced conductance when the temperature and applied bias are less than the Kondo temperature, $\TK$, giving a ZBP with width $\sim2k_B\TK/e$. The fact that the ZBP width remains constant, from $G<0.7\times 2\cond$ all the way down to $10^{-4}\cond$, then implies that $\TK$ is not affected by the overall conductance.  But $\TK$ scales exponentially with the coupling of the localized state to the leads, so it is difficult to explain the insensitivity of $\TK$ to QPC conductance over three orders of magnitude.\cite{HaldanePRL78, GoldhaberPRL98}

A way around this seeming inconsistency, and still within the framework of Kondo physics, could be that localized states in QPCs with low conductance are coupled to leads through strongly asymmetric barriers.\cite{Hong10}  This scenario is supported by density functional calculations, which predict asymmetrically-bound localized states on either end of a QPC near pinchoff (Fig.~4b).\cite{RejecNature06}  The localized states are separated by an opaque barrier that would limit the overall conductance, but each connected to the reservoir through a transparent barrier that sets $\TK$. \cite{GlazmanJPCM04} Since $G$ and $\TK$ are determined by different barriers, their dependence on $\Vg$ could, in principle, be different.  One piece of experimental evidence supporting the asymmetric Kondo model is that ZBPs at low conductance are often observed to be somewhat asymmetric, and not to be centered at exactly zero bias; similar features have been observed in quantum dots with asymmetric contacts.\cite{SimmelPRL99, WysokinskiPRB02}

A classic signature of Kondo-related zero-bias peaks is that they split by $2E_z$ in magnetic field: this has been referred to as the ``smoking gun" of Kondo effect.\cite{MeirPRL93}  Despite the frequent observation of splitting in a magnetic field, in the present experiment and others,\cite{CronenwettPRL02, SarkozyPRB09, ChenPRB09} this smoking gun is less than convincing because the expected splitting magnitude, $2E_z$, is observed in neither high- nor low-conductance regimes.  By comparing the actual peak splitting to the expected $2E_z$, an effective g-factor greater than $0.44$ is typically observed for $G\gtrsim1\cond$, and lower than $0.44$ for $G\ll1\cond$.


The enhanced g-factor observed in the high-conductance regime ($G\gtrsim1\cond$) has often been interpreted as splitting of a Kondo-related ZBP with the exchange-enhanced g-factor that defines subband splittings in QPCs and low-density 2DEGs.\cite{CronenwettPRL02, ThomasPRL96} This reasoning cannot help to explain the reduced g-factor in the low-conductance regime. One explanation for peak splitting less than $2E_z$ would be if the 2DEG wavefunction penetrated significantly into the AlGaAs layer, but this effect is expected to be significant only when electron density is high or the 2DEG is close to the surface,\cite{KoganPRL04} neither of which are the case in this experiment. Alternatively, theoretical calculations that consider details of the peak shape at finite bias predict a somewhat reduced peak splitting $\Dpp\sim4/3E_z$ when $E_z$ is on the order of $\kB\TK$ and $\Dpp\sim1.7E_z$ even at $E_z\sim100\kB\TK$,\cite{MoorePRL00} but this is still much larger than the peak splitting observed in some ZBPs (e.g., Fig.~2d).

Several other theoretical models have been proposed for electron transport in low-density 1D quantum wires, including spontaneous spin polarization, electron-phonon scattering, and Wigner crystalization.  To our knowledge, spontaneous spin polarization would not explain zero-bias conductance peaks.\cite{ThomasPRL96, BerggrenPRB96, ReillyPRL02}  Electron-phonon scattering would give rise to a ZBP, but only a weak magnetic field dependence is expected.\cite{MatveevPRL03}  Partial wigner crystallization in the QPC may suppress the conductance,\cite{MatveevPRL04} and an analog of the Pomeranchuk effect could then explain magnetic field-dependent ZBPs,\cite{SpivakAP06} but this effect is not predicted to give ZBP splitting at finite field.

Recently, it has been suggested that sharp ZBPs could be reproduced with the saddle-point model\cite{ButtikerPRB90, MartinMorenoJPCM92} by taking into account the rise of subband energy with increasing source-drain d.c.~bias $\Vdc$.\cite{ChenPRB09} External bias is usually assumed to drop linearly across the QPC, so the bottom of subband stays fixed with respect to the center of bias window.\cite{MartinMorenoJPCM92, PatelPRB91} At low conductance, however, the linearity of the potential drop is modified by electron-electron interactions. To minimize the interaction energy, the bottom of subband deviates from the center of bias window and has been observed to move upwards,\cite{PicciottoPRL04} consistent with numerical calculations employing a nonequilibrium Green's function formalism.\cite{ZozoulenkoPRB09}

By including a simple linear rise of subband energy with $\Vdc$,\cite{ChenPRB09} the saddle-point model gives ZBPs persisting down to the low-conductance limit with constant FWHM and $\delta G/\gmax$ (Fig.~4c), agreeing remarkably with the experimental results in Fig.~1b.  This explanation is also consistent with the lack of explicit temperature dependence on the barrier, as seen in Fig.~3.  But there have been no predictions, to our knowledge, for a spin dependence of this effect, conflicting with the universal disappearance of the ZBP for large in-plane field. Indeed, the spin degree of freedom is absent from the calculations in Ref.~\onlinecite{ZozoulenkoPRB09}, suggesting that subband energies should rise even for spin-polarized carriers.

\section{Conclusions}
Zero bias conductance peaks in QPCs persist down to $10^{-4}\cond$, but the low- and high-conductance phenomenology is quantitatively and qualitatively different. Three characteristics of low-conductance ZBPs are reported: (1) the FWHMs and ratio $\delta G/\gmax$ does not change with conductance; (2) ZBPs are always suppressed and show some type of splitting in magnetic field, but the splitting is suppressed compared to expectations of Kondo model and often does not appear until the peak has nearly disappeared into the background conductance; and (3) explicit temperature dependence of the QPC transmission coefficient is weaker compared with their high-conductance counterparts. A Kondo model with asymmetric localized states was discussed in connection with these characteristics. Although a FWHM that is independent of conductance can in principle be interpreted in this way, it is difficult to account for the suppression of ZBP splitting. Bias dependence of QPC subband energies can easily explain ZBPs persisting down to pinchoff, but their strong magnetic field dependence is inconsistent with this model.

\acknowledgments{We thank I. Affleck, K. A. Matveev, Y. Meir, B. Spivak, and E. Sela for helpful discussions. Work at UBC was supported by NSERC, CFI, and CIFAR. W.W. acknowledges financial support by the Deutsche Forschungsgemeinschaft (DFG) in the framework of the program ``Halbleiter-Spintronik'' (SPP 1285).}

\bibliographystyle{apsrev}
\bibliography{../mybib}

\end{document}